\documentclass[runningheads]{llncs}
\usepackage{graphicx}
\usepackage{xcolor}

\graphicspath{{fig/}}

\newcommand{\DOI}[1]{\textbf{DOI}: #1}
\begin{document}
\title{Responsible Urban Intelligence: Towards a Research Agenda\thanks{This work was supported in part by the National Natural Science Foundation of China under Grant 42101472 and the Hong Kong Polytechnic University Start-Up under Grant BD41.
}}
\titlerunning{SDSS 2023}
%
%
\author{Rui Cao\inst{1}\orcidID{0000-0002-1440-4175} \and
Qi-Li Gao\inst{2}\orcidID{0000-0003-0179-3500} \and
Guoping Qiu\inst{3}\orcidID{0000-0002-5877-5648}}
\authorrunning{R. Cao et al.}
%
\institute{LSGI \& SCRI, The Hong Kong Polytechnic University, HKSAR, China\\ \email{rucao@polyu.edu.hk} \and
CASA, University College London, UK\\
\email{qili.gao@ucl.ac.uk} \and
School of Computer Science, University of Nottingham, UK\\
\email{guoping.qiu@nottingham.ac.uk}}
\maketitle              
\begin{abstract}
Acceleration of urbanisation is posing great challenges to sustainable development. Growing accessibility to big data and artificial intelligence (AI) technologies have revolutionised many fields and offered great potential for addressing pressing urban problems. However, using these technologies without explicitly considering responsibilities would bring new societal and environmental issues. To maximise the benefits of big data and AI while minimising potential issues, we envisage a conceptual framework of Responsible Urban Intelligence (RUI) and advocate an agenda for action. We first define RUI as consisting of three major components including urban problems, enabling technologies, and responsibilities; then introduce transparency, fairness, and eco-friendliness as the three dimensions of responsibilities which naturally link with the human, space, and time dimensions of cities; and further develop a four-stage implementation framework for responsibilities as consisting of solution design, data preparation, model building, and practical application; and finally present a research agenda for RUI addressing challenging issues including data and model transparency, tension between performance and fairness, and solving urban problems in an eco-friendly manner.

\end{abstract}
\DOI 
https://doi.org/10.25436/E2RC73
\section{Introduction}\label{sec:intro}
\subsection{Urban Problems and Data-driven Solutions}\label{sec:urban}

Urbanisation is one of the most important topics in the 21st century. The majority of the world population now live in cities and two thirds of them are projected to live in urban areas by 2050 \cite{un_wup_2018}. Increasing urban population is posing continuous environmental and social challenges.
To achieve the United Nations' sustainable development goals (SDG) \cite{un_transforming_2015} for cities by 2030, it is urgent to address these problems effectively and efficiently.

Benefiting from the advances in information and communications technology, the growing accessibility to various urban data, including remote sensing data and emerging geospatial big data,
has enabled us to sense cities timely and at an unprecedented level of detail \cite{battyNewScienceCities2013,liuSocialSensingNew2015,caoDeepLearningbasedRemote2020,burkeUsingSatelliteImagery2021,caoIntegratingSatelliteStreetlevel2023}.
The growing data abundance has further promoted the development of data-driven methods. Particularly, recent breakthroughs in Artificial Intelligence (AI) have significantly impacted the fields of geospatial and urban studies, and led to new directions such as GeoAI \cite{janowiczGeoAISpatiallyExplicit2020}, Urban Computing \cite{zhengUrbanComputingConcepts2014}, and Urban Informatics \cite{shiUrbanInformatics2021}.
These enabling technologies offer potential solutions to 
many urban issues \cite{shiUrbanInformatics2021,zhengUrbanComputingConcepts2014,panUrbanBigData2016}.
However, misuse of the technologies and data would also lead to new problems, such as nontransparent decision-making, socio-spatial inequalities, and unsustainable energy consumption \cite{singletonUrbanGovernance2021,gaoSegregationIntegrationExploring2021,strubellEnergyPolicyConsiderations2020}.

\subsection{Responsible AI}\label{sec:rai}
Last decade has witnessed the renaissance of AI fuelled by the advancement of deep learning, which has achieved breakthroughs in many fields \cite{lecunDeepLearning2015}.
Although AI models are becoming more powerful, it can also create potential problems.
On the one hand, current AI models can achieve superior performances on some specific tasks like perception and generation. The misuse of AI, particular AI-generated content (AIGC) such as DeepFake \cite{verdolivaMediaForensicsDeepFakes2020} and the recent ChatGPT \cite{sison2023chatgpt}, is becoming more problematic.
On the other hand, current data-driven AI models have non-negligible flaws such as gender and racial bias \cite{gebruDatasheetsDatasets2021}, excessive energy consumption \cite{strubellEnergyPolicyConsiderations2020} and being black-boxes \cite{barredoarrietaExplainableArtificialIntelligence2020}. 

Recently, increasing attention has been paid to \textit{Responsible AI}, to study how to use AI responsibly.
The field is relatively new and there are different concepts and frameworks.
Dignum \cite{dignumResponsibleArtificialIntelligence2019} proposes the ART principles, including accountability, responsibility, and transparency, as a framework to develop and use AI in a responsible way. 
%
Besides academia, in industry, many technology giants have also proposed their own principles for responsible AI, such as Microsoft\footnote{\url{https://www.microsoft.com/ai/responsible-ai}}, Google\footnote{\url{https://ai.google/responsibilities/responsible-ai-practices}}, and Facebook\footnote{\url{https://ai.facebook.com/blog/facebooks-five-pillars-of-responsible-ai}}.
Their principles differ but share similar ingredients, such as fairness, safety, privacy, inclusiveness, transparency, and accountability.

Currently, data-driven technologies have been widely applied to deal with urban problems, but without explicitly considering how to use them responsibly. To bridge the gap, we advocate for a research agenda of \textit{Responsible Urban Intelligence} (RUI) by explicitly proposing principles of responsibilities in using data-driven technologies to tackle urban problems.

\section{Responsible Urban Intelligence} \label{sec:rui}
Cities are integrated complexes of human, space, and time as well as the interactions among them \cite{battyNewScienceCities2013}.
To achieve urban SDGs \cite{un_transforming_2015} and respond to Responsible AI \cite{dignumResponsibleArtificialIntelligence2019}, 
We herein propose the conceptual framework of RUI 
to explicitly bring the principles of responsibility into the urban context, and advocate for responsible use of data-driven technologies in tackling urban problems and gaining trust from citizens at large.

\subsection{Conceptual Framework}
\subsubsection{Concept}
\textit{Responsible Urban Intelligence} refers to the research and application that use data-driven methods and technologies responsibly to address urban problems for the general good. The ultimate goal is to 
help build human-centred smart cities for the benefit of society,
and eventually achieve sustainable and equitable urban development.

\begin{figure}[h]
	\centering
	\includegraphics[width=0.9\linewidth]{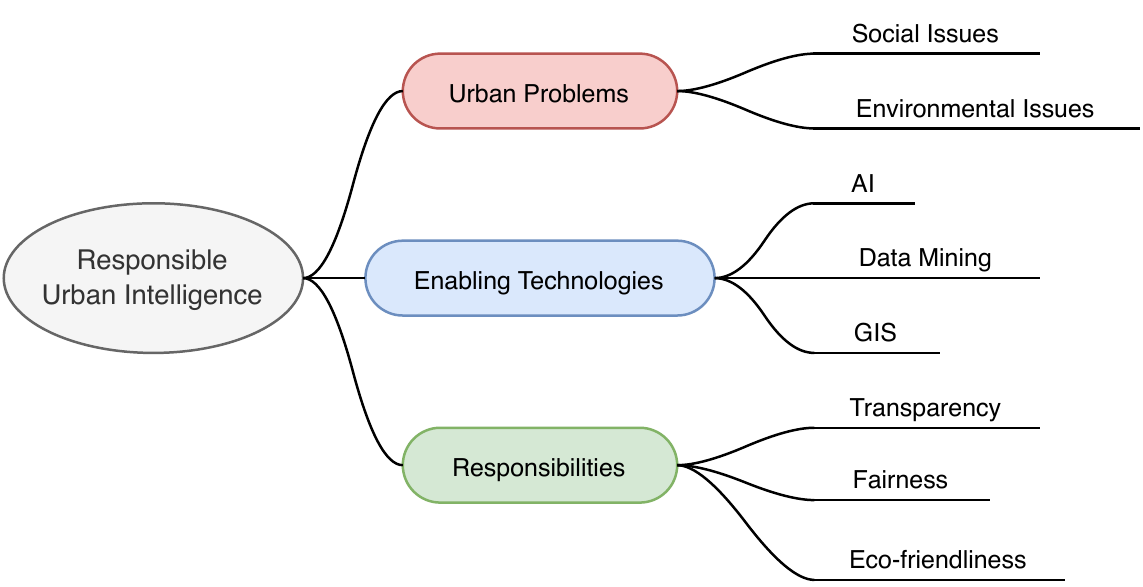}
	\caption{Components of RUI: urban problems, enabling technologies, and responsibilities.}
	\label{fig:rui-components}
\end{figure}   

\subsubsection{Components}
RUI includes three core components as shown in Figure \ref{fig:rui-components}: urban problems, enabling technologies, and responsibilities. 
The \textbf{urban problems} component pinpoints the context and scope that the RUI research deals with, i.e., the significant and urgent urban problems that face the lives of the major world population \cite{un_transforming_2015}. For example, social issues like housing and transport inequalities, and environmental issues such as air pollution and natural hazards. 
%
The \textbf{enabling technologies} component indicates the digital technologies (including AI, data mining, GIS, etc.) that can be exploited to improve the ability to sense and manage cities and tackle urban problems.
With the growing popularity of urban big data, enabling technologies is becoming increasingly important in supporting urban development.
The final component is \textbf{responsibilities}, which we argue is of equal importance to the urban problems per se. However, it is often overlooked in current data-driven solutions to urban problems \cite{shiUrbanInformatics2021}.
This component can be understood from two different perspectives. First, it means that the initial purpose and ultimate goal of the research should be responsible and towards general good. Second, the goals should be achieved in a responsible manner to ensure that the process is transparent, fair, and eco-friendly and does not exacerbate existing issues or create new ones. 

\begin{figure}[h]
	\centering
	\includegraphics[width=0.75\linewidth]{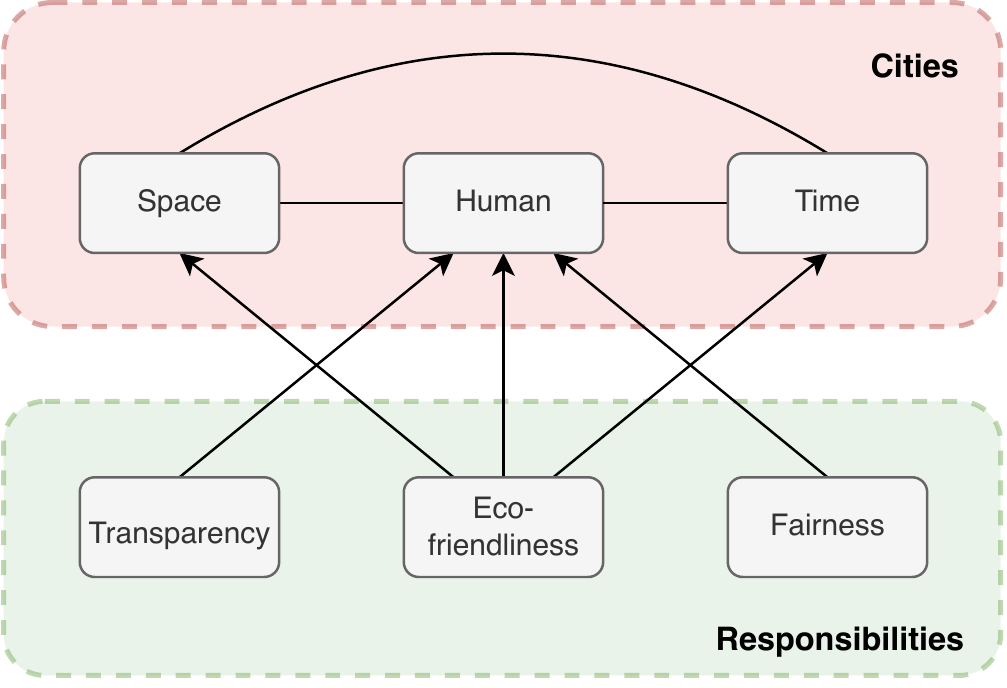}
	\caption{The correspondence of the dimensions of responsibilities to the dimensions of cities.}
	\label{fig:cities-responsibilities-correspondence}
\end{figure}

\subsubsection{Dimensions of Responsibilities}
The component of responsibilities can be decomposed into three fundamental dimensions: transparency, fairness, and eco-friendliness (see Figure \ref{fig:rui-components}). The three dimensions can well correspond to the human-space-time dimensions of cities as shown in Figure \ref{fig:cities-responsibilities-correspondence}.
\textbf{Transparency} requires the algorithmic decision-making to be transparent to both decision makers and the civic public. This is critical for human-centred urban governance and development \cite{singletonUrbanGovernance2021}.
In fact, transparency inherently demands for the explainability of models and algorithms, which resonates the current needs for unveiling the black-box data-driven models \cite{barredoarrietaExplainableArtificialIntelligence2020}.
It is also the essential basis for human demands on security, privacy, and reliability \cite{dignumResponsibleArtificialIntelligence2019}.
\textbf{Fairness} requires that the solution and model include enough diversity and inclusiveness to reduce the likelihood of incurring new injustices or worsening the current situation when addressing urban problems (e.g., urban housing and transport inequalities \cite{gaoSegregationIntegrationExploring2021}).
It is a critical part to realise the shift of focus from purely pursuing performance and efficiency to accounting for justice and humanities. This is crucial for human-centred solutions to urban problems.
\textbf{Eco-friendliness} is another important dimension that correlates with all the human-space-time dimensions of cities. It demands explicit consideration of sustainable development from an environmental perspective in the solutions to urban problems.
For example, the training of large deep learning models can consume noticeable energy and result in non-negligible carbon emission, so it is important to take eco-friendly measures to alleviate this situation and prevent diminishing efforts in tackling global warming \cite{strubellEnergyPolicyConsiderations2020}. 

\subsection{Towards a Research Agenda}
\subsubsection{Implementation Framework}
\begin{figure}[h]
	\centering
	\includegraphics[width=0.8\linewidth]{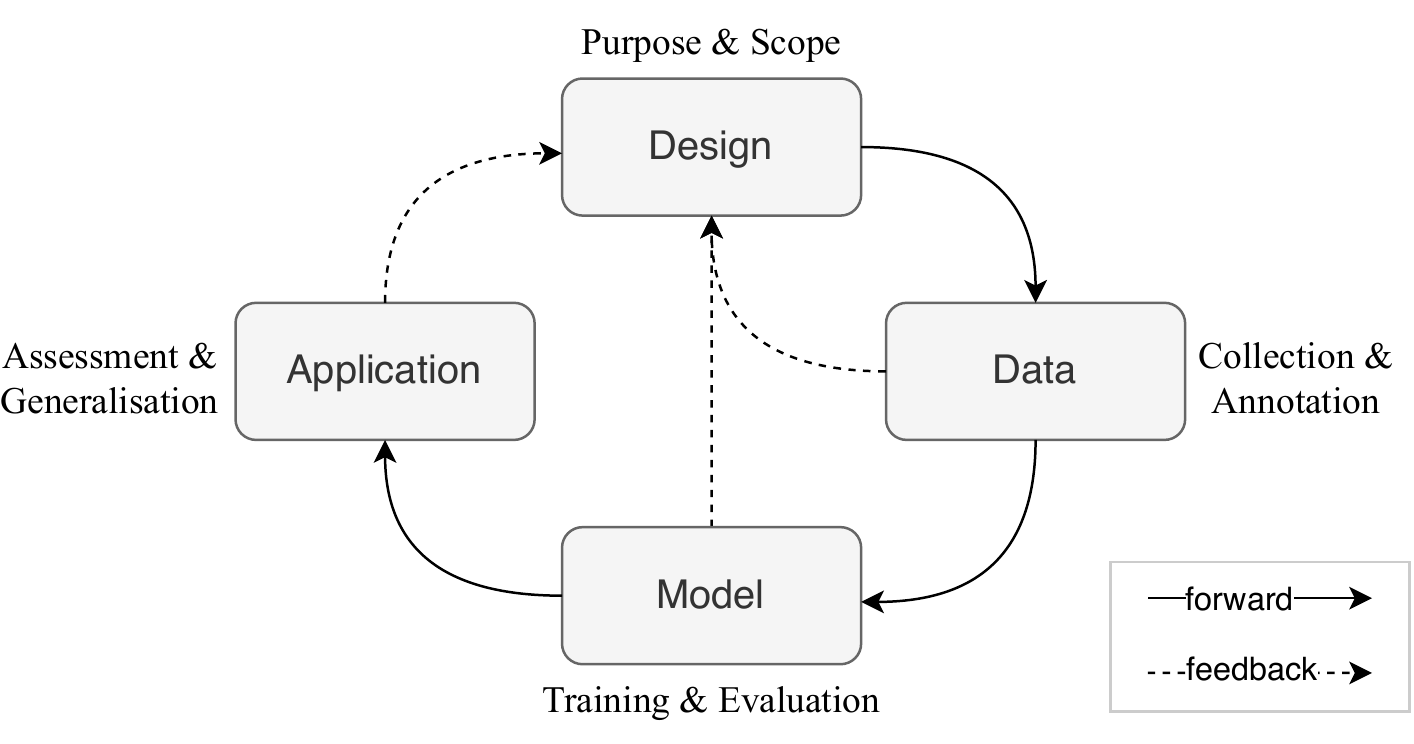}
	\caption{The implementation framework for responsibilities, which can be applied in different stages.}
	\label{fig:rui-framework}
\end{figure}

The principles of responsibilities can be applied in different stages of the solutions to urban problems, 
as shown in Figure \ref{fig:rui-framework}.

In the \textbf{solution design} stage, the responsibilities can be reflected on the initial purpose and scope of the solution, which should be designed for the general good and remain responsible for stakeholders. Meanwhile, the design process should account for the principles of transparency, fairness, and eco-friendliness. 
%
In \textbf{data preparation}, the principles of transparency and fairness should be highlighted.
As the quality of datasets is critical for building data-driven AI models and errors and biases in datasets would be directly inherited by the models trained upon them \cite{gebruDatasheetsDatasets2021}, 
data preparation procedures such as data collection, preprocessing, and annotation should be kept transparent and fair to increase model reliability.
In \textbf{model building}, both the transparency, fairness, and eco-friendliness principles can be applied.
The used models should pursue explainability from different levels \cite{barredoarrietaExplainableArtificialIntelligence2020}.
The fairness factors such as inclusiveness and diversity should be explicitly considered to avoid biases and inequalities \cite{gebruDatasheetsDatasets2021}.
In addition, models should aim for simplicity and efficient training to support eco-friendly computation and energy consumption \cite{strubellEnergyPolicyConsiderations2020}.
In the \textbf{practical application} stage, the dimensions of responsibilities can also guide the practical use. The models and algorithms should stick to the initial design purpose and scope, and be enforced with transparency, fairness, and eco-friendliness goals for assessment. The generalisation of the solutions to problems beyond the initial scope should also be treated responsibly and with discretion.

The entire process of the stages is dynamic and iterative, and the feedback from the application stage can be exploited to further improve the solution design in return.
Furthermore, the intermediate feedback from the data collection and model building stages can also be used to refine the solution design in a timely manner.

\subsubsection{Challenges and Agenda}
To achieve aforementioned responsible design and use of data-driven solutions to urban problems, there are still many challenges. An research agenda to tackle the pressing ones is advocated.

A first area of the research agenda is to address the pressing needs for data and model transparency.
Open urban data is critical for transparency, yet their quantity and quality are now insufficient and diverse, and their effective use remains questionable \cite{singletonUrbanGovernance2021}. Another challenge is the data privacy facing the open data.
Additionally, the explainability and reasoning of models are crucial for many applications, especially for high-stakes scenarios.
Currently, many accurate data-driven models are difficult to interpret and only indicate correlations rather than causality, which hinders their credibility \cite{barredoarrietaExplainableArtificialIntelligence2020}.
How to reconcile the contradiction between accuracy and transparency deserves further efforts.

The second area is to relieve the tension between performance and fairness.
Most enabling technologies are designed to improve performance and efficiency, while it remains challenging to accommodate fairness, especially as there is no uniform standard \cite{dignumResponsibleArtificialIntelligence2019}. It is thus difficult to take fairness into account and pay more attention to inclusion, diversity, and equity.
In addition, civic participation is one of the central elements for human-centred urban development, and fairness and justice cannot be easily achieved without a full representation of different social groups \cite{shiUrbanInformatics2021}. Therefore, how to engage urban residents in the research and application processes, especially those underrepresented, is also a challenge.

A third area of research is to study how digital technologies can be adapted to urban scenarios while achieving eco-friendly goals.
Current AI models, particularly deep learning models, are becoming increasingly large and their training will cost significant amounts of energy and cause outstanding carbon emissions \cite{strubellEnergyPolicyConsiderations2020}. 
To pursue eco-friendliness, there are two main challenges. The first is to reduce the need for large models. We can opt for more lightweight models or build models on top of pretrained ones depending on specific scenarios.
The other is to enhance top-down design and planning to create domain-specific computing infrastructures for urban studies, and also promote the sharing of data and models.
This can help reduce the redundant computing and also increase efficiency.

The research of RUI is highly interdisciplinary and requires efforts from a variety of fields, including urban studies, AI, GIScience, social science, environmental science, etc. We thus call for coordinated efforts to meet the challenges and work towards the common SDGs for cities in a responsible way.

\section{Conclusion}\label{sec:conclusion}
Urbanisation is one of the most important topics of the 21st century, and urban problems are becoming increasingly prominent. To tackle these challenges, data-driven technologies have been widely used today. However, without the guidance of responsibilities, existing issues can be exacerbated and new problems can be introduced.
To minimise the negative effects of leveraging these technologies in addressing urban issues, the principles of responsibilities should be clearly established and emphasised.
To this end, we envisage the conceptual framework of RUI and advocate the research agenda as a feasible pathway towards United Nations' SDGs for cities.
This agenda will greatly increase the awareness of responsibilities and will also promote the sustainability of using enabling technologies to solve urban problems.

%
%
%
\bibliographystyle{splncs04}
\bibliography{ref,myref}

\end{document}